\documentclass[twocolumn,aps,pre,floatfix, superscriptaddress, showpacs]{revtex4-1}

\usepackage{amsmath}
\usepackage{graphicx}
\usepackage{hyperref}
\usepackage[caption=false]{subfig}

\newcommand{\dd}[2]{\frac{d#1}{d#2}}
\newcommand{\critpoint}{0.437344654(21)}

\begin{document}

\title{You Can Run, You Can Hide:\\
The Epidemiology and Statistical Mechanics of Zombies}
\author{Alexander A. Alemi}
\email{aaa244@cornell.edu}
\affiliation{Laboratory of Atomic and Solid State Physics, Cornell University, Ithaca, NY 14853}
\author{Matthew Bierbaum}
\email{mkb72@cornell.edu}
\affiliation{Laboratory of Atomic and Solid State Physics, Cornell University, Ithaca, NY 14853}
\author{Christopher R. Myers}
\email{c.myers@cornell.edu}
\affiliation{Laboratory of Atomic and Solid State Physics, Cornell University, Ithaca, NY 14853}
\affiliation{Institute of Biotechnology, Cornell University, Ithaca, New York}
\author{James P. Sethna}
\email{sethna@lassp.cornell.edu}
\affiliation{Laboratory of Atomic and Solid State Physics, Cornell University, Ithaca, NY 14853}
\date{\today}


\begin{abstract}
We use a popular fictional disease, zombies, in order to introduce techniques
used in modern epidemiology modelling, and ideas and techniques used in the
numerical study of critical phenomena.  We consider variants of zombie models,
from fully connected continuous time dynamics to a full scale exact stochastic
dynamic simulation of a zombie outbreak on the continental United States. Along
the way, we offer a closed form analytical expression for the fully connected
differential equation, and demonstrate that the single person per site two
dimensional square lattice version of zombies lies in the percolation
universality class.  We end with a quantitative study of the full scale US
outbreak, including the average susceptibility of different geographical
regions.
\end{abstract}

\maketitle

\section{Introduction}

Zombies captivate the imagination.  The idea of a deadly disease that not only
kills its hosts, but turns those hosts into deadly vectors for the disease is
scary enough to fuel an entire genre of horror stories and films.  But at its
root, zombism is just that -- a (fictional) disease -- and so should be amenable to the same kind
of analysis and study that we use to combat more traditional diseases.

Much scholarly attention has focused on more traditional human diseases
\cite{book}, but recently, academic attention has turned a bit of thought
onto zombies as a unique and interesting modification of classic disease
models. One of the first academic accounts of zombies was the 2009 article by
Munz et al.~\cite{smith}, in which an early form of a compartmental model of
zombism was introduced. Since then, there have been several interesting papers
published including works that perform Bayesian estimations of the zombie
disease parameters~\cite{bayesian}, look at how emotional factors impact the
spread of zombies~\cite{rulezombies}, using zombies to gain insight into models
of politics~\cite{hochreiter2014zombie}, or into the interaction of a zombie
epidemic and social dynamics~\cite{sander2014zombie, poster}.  Additional essays can be
found in two books collecting academic essays centered around zombism~\cite{zombiebook1, zombiebook2}

Besides the academic papers, zombies have seen a resurgence in
fiction.  Of particular note are the works of Max Brooks, including a 
detailed \emph{Zombie Survival Guide}~\cite{survivalguide}, as well as an oral
history of the first zombie war~\cite{worldwarz} in a hypothesized post
outbreak world. In both these works Brooks provides a rich source of
information about zombies and their behavior.  In particular,
he makes the connection to disease explicit, describing zombies as the result
of a hypothetical virus, \emph{Solanum}.

Zombies form a wonderful model system to illustrate modern epidemiological
tools drawn from statistical mechanics, computational chemistry, and
mathematical modeling. They also form an ideal vehicle for public outreach:
the Center for Disease Control uses preparation for a zombie
apocalypse~\cite{url:CDCZombies1,url:CDCZombies2} to promote emergency
preparedness.  In this work, we will build up to a full-scale
simulation of a zombie outbreak in the continental United States, with 
realistic values drawn from the literature and popular culture
(section~\ref{sec:US}, simulation accessible online \footnote{\url{http://mattbierbaum.github.io/zombies-usa/}}).
Before that, we shall use statistical mechanics to scrutinize the threshold
of zombie virulence that determines whether humanity survives
(section~\ref{sec:RG}). Preceding that, we shall show how methods
from computational chemistry can be used to simulate every individual heroic
encounter between a human and a zombie (section~\ref{sec:Gillespie}). But
we begin by describing and analyzing a simple model of zombies (the $SZR$ 
model) -- the simplest and most natural generalization to the classic
$SIR$ (Susceptible-Infected-Recovered) model used to describe infectious disease spread in epidemiology.

\section{$SZR$ Model}
\label{sec:SZR}

We start with a simple model of zombies, the $SZR$ model. There are three
compartments in the model: $S$ represents the susceptible population, 
the uninfected humans; $Z$ represents the infected state, 
zombies; and $R$ represents our removed state, in this case zombies that have
been terminated by humans (canonically by destroying their brain so as to
render them inoperable).  There are two transitions possible: a human can
become infected if they are bitten by a zombie, and a zombie can be destroyed
by direct action by a human.  There are two parameters governing these
transitions: $\beta$, the bite parameter determines the probability by which a
zombie will bite a human if they are in contact, and $\kappa$ the kill
parameter that gives the probability that a human kills the zombie.  Rendered
as a system of coupled differential equations, we obtain, for a particular
interaction site:
\begin{align} 
    \dot S &= -\beta S Z \\ 
    \dot Z &= (\beta - \kappa) SZ \\
    \dot R &= \kappa SZ
\end{align}
Notice that these interactions are \emph{density dependent}, in the sense that
the rate at which we convert humans to zombies and kill zombies is dependent on
the total count of zombies and humans in this site.  This is in contrast with
most models of human diseases, which frequently adopt \emph{frequency
dependent} interactions wherein $S,Z,R$ would have been interpreted as
the fraction of the population in the corresponding state.

This distinction will become stark once we consider large simulations with very
inhomogeneous populations.  By claiming that zombies can be modeled by a single
bite parameter $\beta$ that itself is a rate per person per unit time, we are
claiming that a zombie in a block with 5,000 people would be one hundred times
as effective at infecting new zombies as a zombie in a block with fifty people;
similarly the zombie in question would be killed one hundred times faster. This
would seem false for an ordinary disease like the flu, but in the case of
zombies, we argue that it is appropriate. Zombies directly seek out hosts to
infect, at which point the human and zombie engage in a duel to the (un)death.

To facilitate analysis we can nondimensionalize the equations by choosing a relevant
population size $N$, and recasting in terms of the dimensionless time parameter
$\tau = t\beta N$ and dimensionless virulence $\alpha = \kappa / \beta $
\begin{align} 
\label{eq:SZR}
    \dd{S}{\tau} &= -\frac{SZ}{N} \nonumber\\
    \dd{Z}{\tau} &= (1-\alpha) \frac{SZ}{N} \\
    \dd{R}{\tau} &= \alpha \frac{SZ}{N} \nonumber
\end{align}

Unlike a traditional disease (e.g., as modeled by $SIR$), for the zombie model, we have a stable
configuration when either the human or the zombie population is defeated ($S=0$
or $Z=0$).  Furthermore, unlike $SIR$, $SZR$ admits an analytical solution,
assuming $R(0) = 0$, and with $Z_0 \equiv Z(0), S_0 \equiv S(0)$:
\begin{align} 
    P &\equiv Z_0 + (1-\alpha) S_0 \\
    \mu &\equiv \frac{S_0}{Z_0} (1 - \alpha) = \frac{P}{Z_0} - 1 \\ 
    f(\tau) &\equiv \frac{P \mu }{ e^{\tau P/N } +
\mu } \\ 
    Z(\tau) &= P - f(\tau) \\ S(\tau) &= \frac{f(\tau)}{1-\alpha}
\end{align}
Given the analytical solution, it is clear to see that the sign of $P$ governs
whether there will eventually be humans or zombies in the final state. If
$\alpha < 1, P > 0$, so 
\begin{align}
    \lim_{\tau \to \infty} f(\tau) &= 0 \\
    \lim_{\tau \to \infty} Z(\tau) &= P = Z_0 + (1-\alpha) S_0 \\
    \lim_{\tau \to \infty} S(\tau) &= 0 
\end{align}
and the system will always flow to a final state composed of entirely zombies and no
humans, where $P$ denotes the number of zombies that survive.

If however, $\alpha > 1$, humans are more effective at killing zombies
than zombies are at biting humans. With enough zombies in the initial state,
we can still convert all of the humans before they have time to kill
all of the zombies.

We can recast the dynamics in terms of the variables $P \equiv Z +
(1-\alpha)S$ and $\chi = S/Z$ to gain further insights.  First note that:
\begin{align}
    \dd{P}{\tau} &= P' = Z' + (1-\alpha) S' \\
                 &= (1-\alpha)\frac{SZ}{N} - (1-\alpha)\frac{SZ}{N} = 0
\end{align}
so $P$ is a constant of the dynamics. As for $\chi$:
\begin{align}
    \chi' &= \frac{S'}{Z} - \frac{SZ'}{Z^2} \\
          &= -\frac{S}{N} - (1-\alpha) \frac{S}{N} \frac{S}{Z} \\
        &= -\frac{S}{N} \left( 1 + (1-\alpha) \right) \chi \\
        &= -\frac{P}{N} \chi 
\end{align}
Hence if we choose $N = |P|$, we end up with the very simple dynamics:
\begin{align}
    P'(\tau) &= 0 \\
    P(\tau) &= P_0 = Z(\tau) + (1-\alpha)S(\tau) = Z_0 + (1-\alpha) S_0 \\
    \chi'(\tau) &= \begin{cases} -\chi & P>0 \\ +\chi & P < 0 \end{cases} \\
    \chi(\tau) &= \frac{S(\tau)}{Z(\tau)} = \chi_0 \begin{cases}  e^{-\tau} & P > 0 \\ e^{+\tau} & P < 0 \end{cases} \\
    \chi_0 &\equiv \frac{S_0}{Z_0} 
\end{align}
Here we see that the dynamics is simply an exponential decay or increase in the
ratio of humans to zombies $\chi = S/Z$. The final populations in either case are
easy to see due to the conservation of $P$.  If zombies win we have
\begin{equation}
    Z_\infty = Z_0 + (1-\alpha)S_0
\end{equation}
And if humans win
\begin{equation}
    S_\infty = S_0 - \frac{Z_0}{\alpha - 1}
\end{equation}

\subsubsection{$SIR$ model}

This dynamics should be compared to the similarly nondimensionlized 
density-dependent $SIR$ model:
\begin{align} \dd{S}{\tau} &= -\frac{SI}{N} \\
\dd{I}{\tau} &= \left(\frac{S}{N} - \mu\right) I \\ \dd{R}{\tau} &= \mu I
\end{align} 
Here $\tau = t\beta N$ as above, but $\mu = \nu / (\beta N) =
R_0^{-1}$, because in the $SIR$ model our infected population recovers on its
own. This is contrasted with $SZR$, where the process of infection and
recovery have the same functional form, depending on the product $SZ$.  This
$\mu$ is the inverse of the usual $R_0$ parameter used to denote the
infectivity of the $SIR$ model, here used to make a closer analogy to the $SZR$
model.  It is this parameter that
principally governs whether we have an outbreak or not.
Unlike the $\alpha$ parameter for $SZR$ which depends only on our
disease constants $\beta, \kappa$, the relevant virulence for the density
dependent $SIR$ model ($\mu$) has a population dependence. 

Notice again that while the only stable configuration for the $SIR$ model is when there 
is no infected population ($I=0$), the $SZR$ model is stable when either the
humans or zombies are depleted ($S=0$ or $Z=0$).  

The $SIR$ model does not admit a closed form analytical solution, but 
we can find a parametric solution by dividing the first equation by the third, revealing.
\begin{equation}
    S(\tau) = S_0 e^{-\frac{(R(\tau) - R_0)}{\mu N} }
\end{equation}
Using the observation that in the limit of infinite time, no infected population can persist,
we can choose $N$ to be the total population
\begin{equation}
    S_0 + I_0 + R_0 = N = S_\infty + R_\infty 
\end{equation}
and so obtain a transcendental equation for the recovered population at long times.
\begin{equation}
    R_\infty = N - S_0 e^{-\frac{(R_\infty - R_0)}{\mu N}}
\end{equation}

Unlike the $SZR$ model, here we see that no matter how virulent the disease is, the epidemic will
be self-limiting, and there will always have some susceptibles left at the end of the outbreak. This is
a sharp qualitative difference between zombies and more traditional $SIR$
models, arising from the fact that the ``recovery'' of zombies is itself dependent on the presence of susceptibles.

\begin{figure}[htbp]
    \begin{center}
    \includegraphics[width=\linewidth]{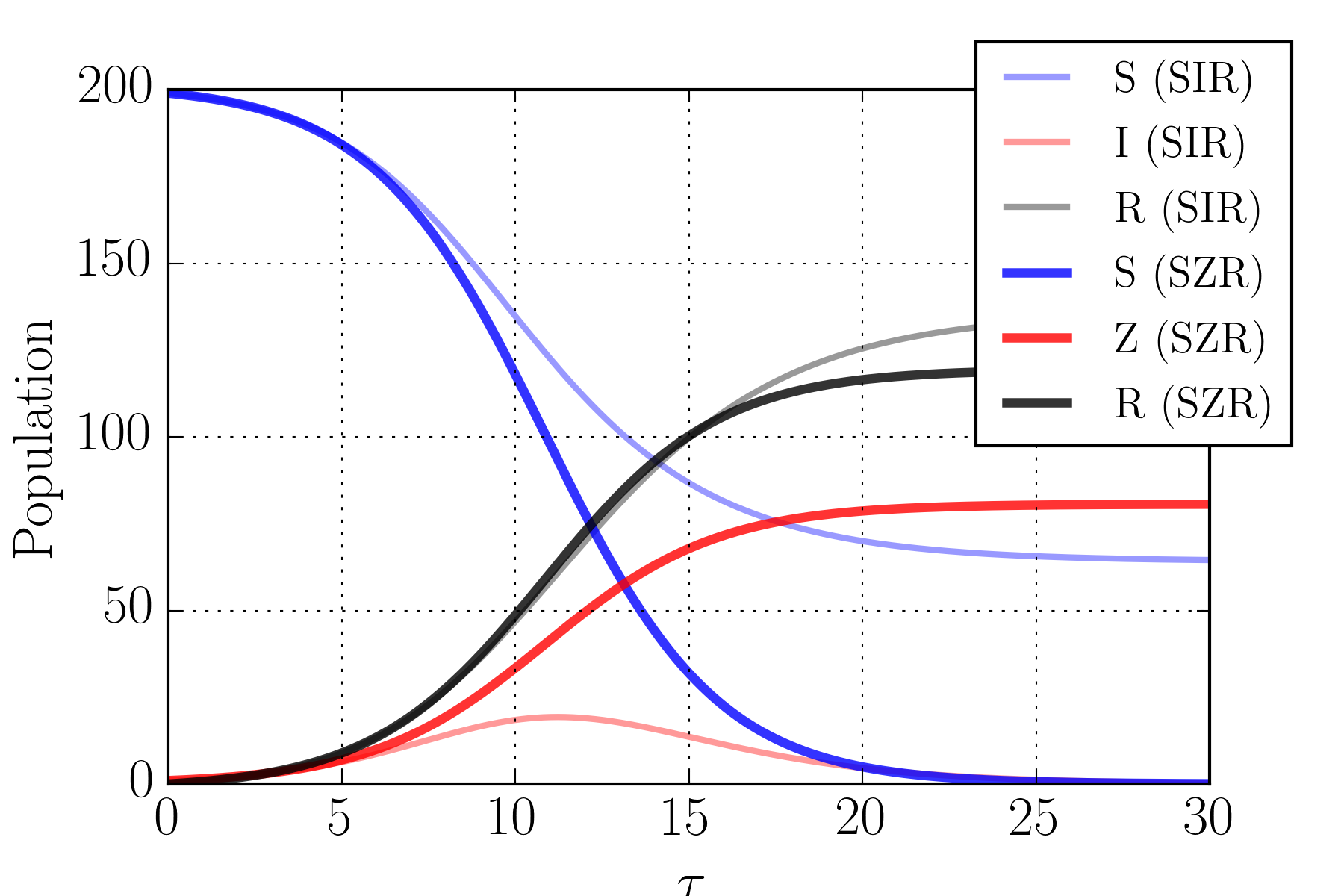}
    \end{center}

    \caption{Deterministic trajectories for the $SIR$ and $SZR$ models with an
        initial population of 200 people, 199 uninfected and 1 infected.  The
        (susceptible, infected, removed) population is shown in (blue, red,
        black) (color online).  The $SZR$ results are solid lines while the $SIR$ results are
        lighter lines. For both models $\tau = t\beta N$ where $N$ was taken to
        be the total population.  For the $SZR$ model $\alpha$ was chosen to be
        0.6, while for the $SIR$ model $\mu$ was chosen to be 0.6 to show
        similar dynamics.  Notice that in this case, in $SZR$ the human
        population disappears and only zombies remain in the end, while
        the $SIR$ model is self-limiting, and only a fraction of the population
        ever becomes infected.} 

\label{fig:analytic_example} 
\end{figure}

To visually compare the difference, in Figure~\ref{fig:analytic_example} we have shown deterministic trajectories
for both $SIR$ and $SZR$ for selected parameter values.

\begin{figure}[htbp] 
    \includegraphics[width=\linewidth]{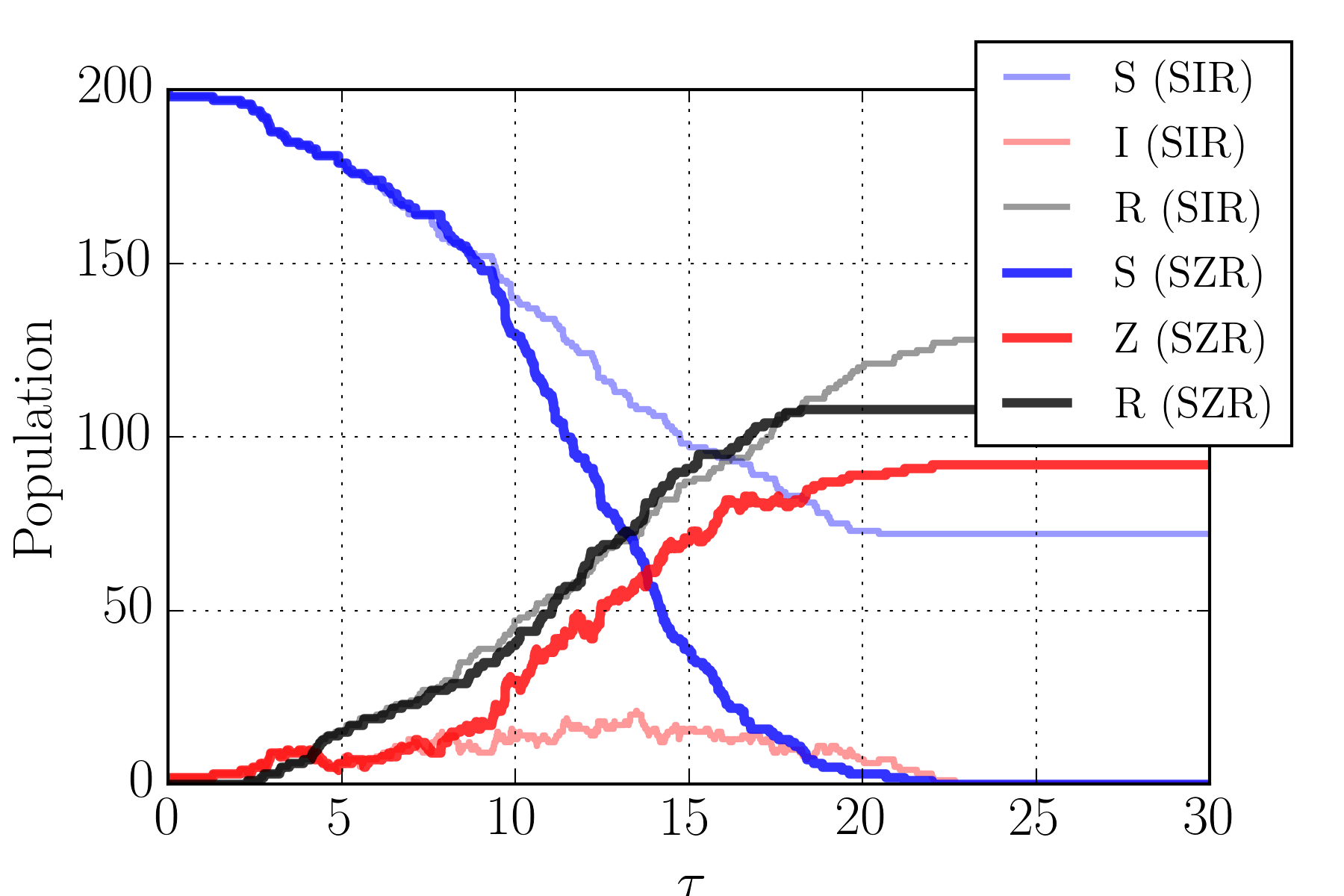}

    \caption{Example Gillespie dynamics for the $SIR$ and $SZR$ models with the
        same parameter settings as Figure~\ref{fig:analytic_example}.  
        The (susceptible, infected, removed) population is shown in (blue, red, black) (color online).  
    The $SZR$ results are solid lines while the $SIR$ results are lighter lines. The two simulations
were run with the same seed so as to match their dynamics at early times.}
    \label{fig:Gillespie_example} 
\end{figure}

\section{Stochastic simulation}
\label{sec:Gillespie}

While most previous studies modeling zombie population dynamics have been 
deterministic, things get more interesting when we try to
model discrete populations. By treating the number of zombies and humans
as continuous variables in the last section, we are ignoring the random
fluctuations that arise in small populations: even a ferociously virulent
zombie infestation might fortuitously be killed early on by happy accident.
Similar problems arise in chemical reactions: reactions involving two types
of proteins in a cell can be described by chemical reaction kinetics evolving
their concentrations (like our $SZR$ equations~\ref{eq:SZR}), but if the
number of such proteins is small, accurate predictions must simulate
the individual binary reactions (each zombie battling each human).
Interpreting our $SZR$ transitions as reaction rates, gives us a system
akin to a chemical reaction with two possible transitions:
\begin{align*}
    (S,Z) \xrightarrow{\beta SZ} (Z,Z) \\
    (S,Z) \xrightarrow{\kappa SZ} (S,R) \\
\end{align*}
When a human and zombie are in contact, the probability of a bite in a small
period of time is given by the bite rate and the size of the populations of the two species
$(\beta S Z \, dt)$, and similarly for the probability of a kill.  In order to
efficiently simulate this dynamics, we use the Gillespie algorithm~\cite{gillespie},
which efficiently uses the computer to sequentially calculate the result
of each one-on-one battle.

The stochasticity gives more character to the simulation. The fully connected
continuous dynamics modeled by the differential equation is straightforward:
either the humans win and kill all of the zombies, or the zombies win and bite
all of the humans.  While the continuous approximation may be appropriate at
intermediate stages of the infection where the total population is large and
there are a non-trivial number of infected individuals, we will eventually be
interested in simulating an actual outbreak on an inhomogeneous population
lattice, where every new site will start with a single infected individual.
But even though we may be interested in modeling the outbreak case ($\alpha <
1$), we would like to allow the possibility that the humans manage to defeat
the outbreak before it really takes off.  The stochastic Gillespie dynamics
allows for this possibility.

In Figure~\ref{fig:Gillespie_example} we have shown an example of a single
stochastic simulation using the same parameter settings as those used in Figure
\ref{fig:analytic_example}.  The stochastic trajectory overall tracks the analytic
result, but at points in the simulation there may be more or fewer zombies than
anticipated if the dice fall that way.

\begin{figure}[htbp] 
    \includegraphics[width=\linewidth]{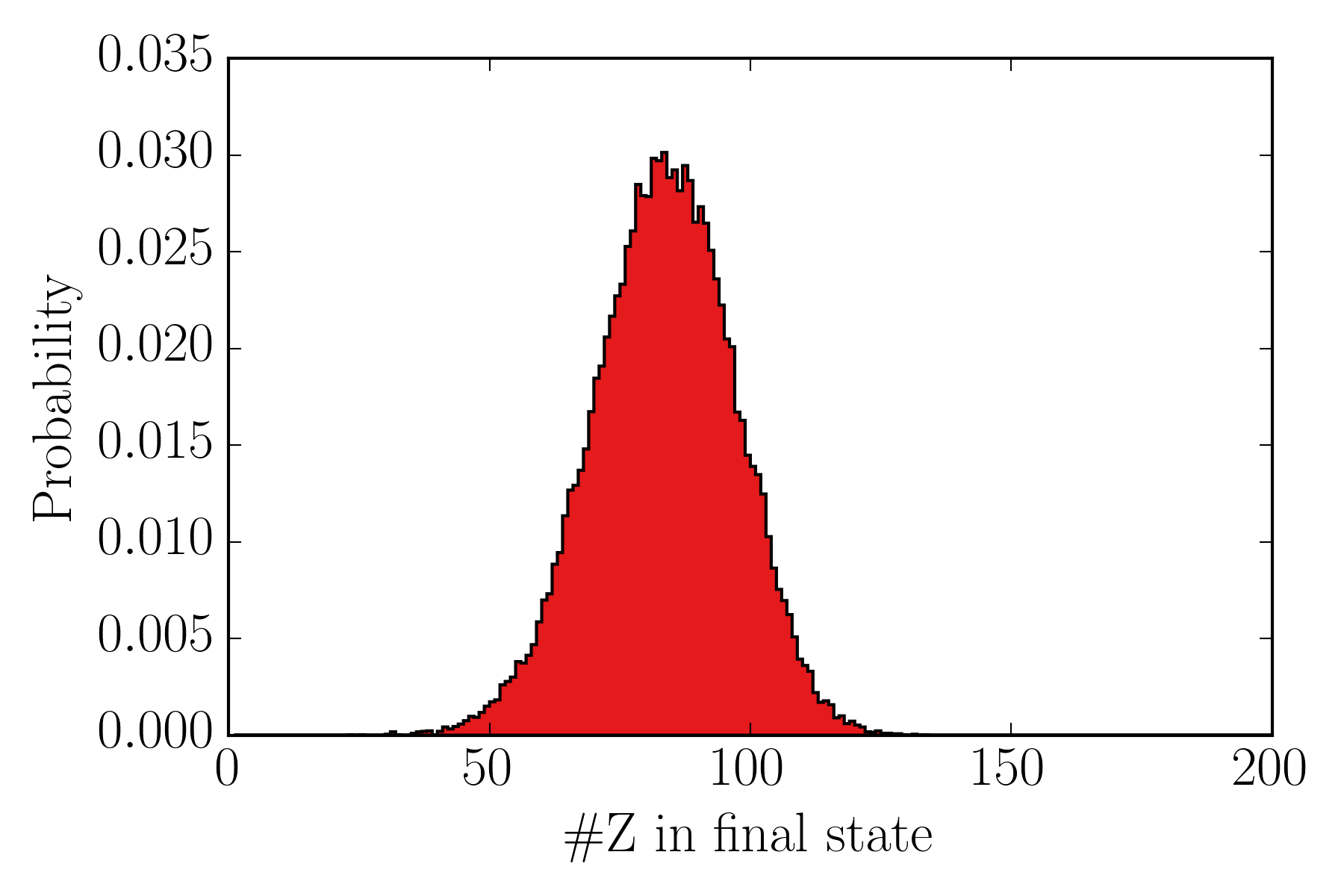}

    \caption{Distribution for final zombies over 100,000 stochastic trajectories with the 
      same parameters as Figure~\ref{fig:Gillespie_example}. Not pictured are
        the 60\% of runs that end with no zombies in the final state. Compare these to the 
    analytical result, in which the final population of zombies would be 81 with
    no possibility of surviving humans. }
    \label{fig:zdist} 
\end{figure}

Another implication of stochastic dynamics is that it is not always guaranteed
that a supercritical ($\alpha < 1$) outbreak will take over the entire susceptible population.
For the parameter settings used in Figure~\ref{fig:analytic_example} and
\ref{fig:Gillespie_example}, namely $\alpha = 0.6$ with a population of 200 and
one infected individual to start, the zombies win only 40\% of the time.
Additionally, the number of zombies we end with is not fixed, as shown in Figure
\ref{fig:zdist}.

\begin{figure}[htbp] 
    \includegraphics[width=\linewidth]{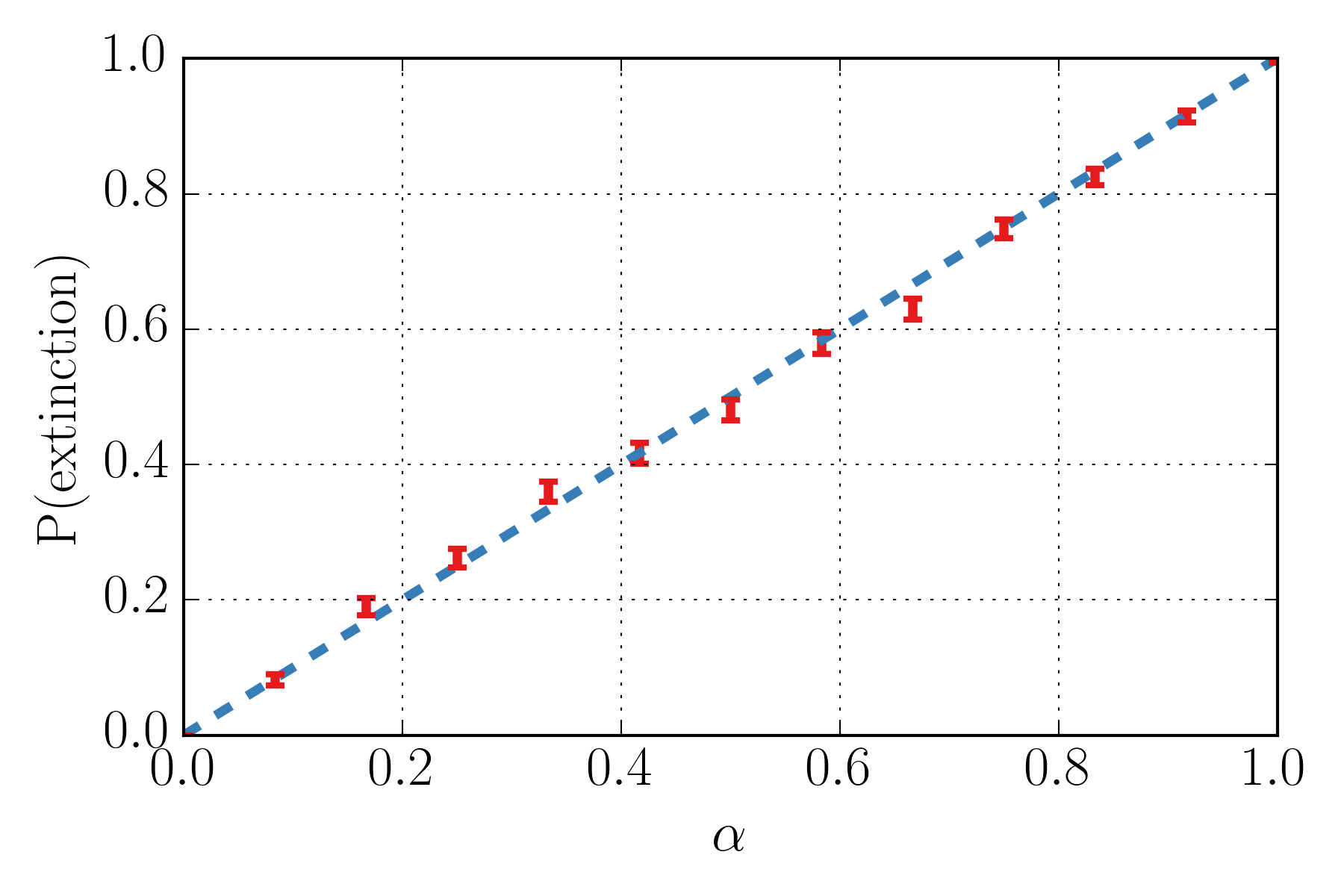}

    \caption{ The observed fraction of simulations that end in an extinction
        for the zombie outbreak, for 1,000 runs of $10^4$ individuals at
        various values of $\alpha$ (eqn.~\ref{eqn:epi}).  The observed extinction probabilities agree
        with the expectation that they should go as $\alpha$, here shown as
    the dashed line. This is the same behavior as the $SIR$ model. }

    \label{fig:epiprob} 
\end{figure}

In fact, we can solve exactly for the probability $P_{\text{ext}}$
that an $\alpha < 1$ simulation will go
extinct in the limit of large populations, using an argument drawn from
the traditional SIR literature. 
At the very beginning of the simulation, there is only one
zombie, who will be killed with probability $\kappa/ (\beta + \kappa)$.  If the first
zombie is killed before it bites anyone, we guarantee extinction.  Otherwise, the zombie will bite another human,
at which point there will be two independent zombie lines that need to be extinguished, which will
occur with probability $P_{\text{ext}}^2$.
This allows us to solve:
\begin{align}
    P_{\text{ext}} &= \frac{\kappa}{\beta + \kappa} 1 + \frac{\beta}{\beta+\kappa} P_{\text{ext}}^2 \\
    P_{\text{ext}} &= \frac{\kappa}{\beta} = \alpha \label{eqn:epi} \ .
\end{align}
The probability of extinction is just given by our dimensionless inverse
virulence $\alpha$.  In Figure~\ref{fig:epiprob} we have shown the observed
extinction probabilities for 1,000 Gillespie runs of a population of $10^4$
individuals at various values of $\alpha$, and overlaid our expected 
dependence of $\alpha$.

This same extinction probability ($P_{\text{ext}} = \mu = R_0^{-1}$) is
observed for the $SIR$ model~\cite{book}.  This is not a coincidence.  
In precisely the limit that is important for studying the probability of an extinction event,
namely at early times with very large populations, the $SZR$ model and $SIR$ are effectively the same,
since the population of susceptibles ($S$) is nearly constant.  Writing $S$ as $S_0 -\delta S$, we have:
\begin{align}
    \frac{dZ}{d\tau} &= (1- \alpha) \frac{S_0 Z}{N} - (1-\alpha) \frac{ (\delta S) Z }{N} \\
    \frac{dI}{d\tau} &= \left( 1 - \frac{\mu N}{S_0} \right) \frac{S_0 I}{N}  - ( \mu N + \delta S) \frac{I}{N}\ .
\end{align}
Here as $\delta S \to 0$, the two models are the same with $\alpha = \mu N / S_0$, another indication that
the density dependent $SIR$ model's virulence is dependent on population size.

\begin{figure}[htbp]
    \includegraphics[width=\linewidth]{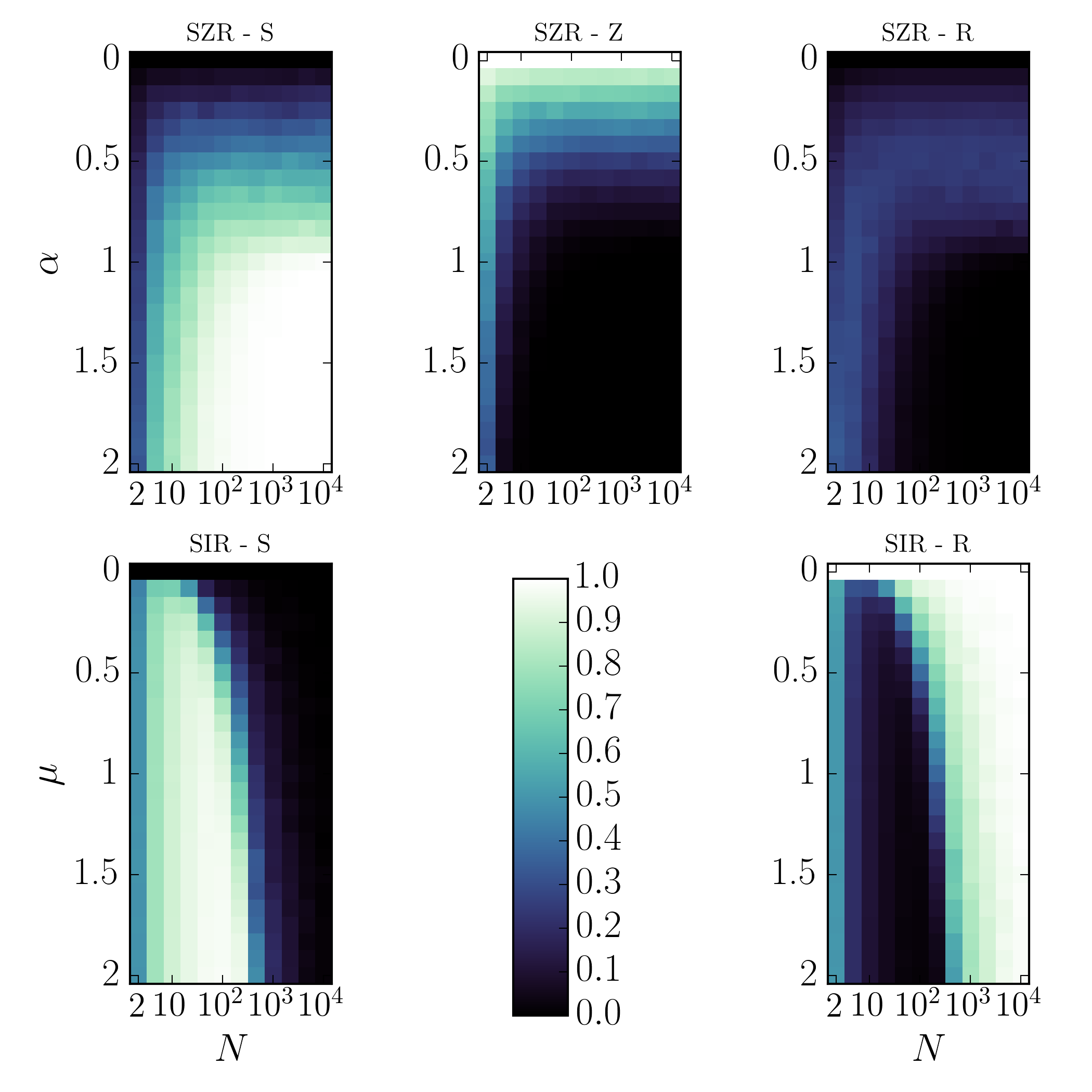}

    \caption{ Mean final states as a function of model parameters.  One thousand different
        simulations are run for each cell.  Each simulation starts with a
        single zombie or infected individual.  The runs are run until they
        naturally terminate, either because the susceptible population is
        deleted, the zombie population is gone, or there are no more infected
        individuals.  Each cell is colored according to the mean fraction of
        the population occurring in each state.  The top row is for $SZR$
        simulations and the bottom row is for $SIR$ simulations.  In both cases
        $N$ is chosen to be 100. Here the sharp contrast between density-dependent 
        $SZR$ and $SIR$ is made apparent. Notice that
        density-dependent $SIR$ is very strongly population dependent.  }

\label{fig:Gillespie_means} 
\end{figure}

To get a better sense of the effect of the stochasticity, we can look at the
mean fractional population in each state for various settings of $\alpha$ and
choices for initial population size.  The results are shown in Figure
\ref{fig:Gillespie_means}.

Plotted are the fractional populations in the final state left for both the $SZR$
model (top row) and $SIR$ model (bottom row) for different parameter combinations
of $\alpha$ and the initial population. In all cases, the $N$ parameter was
chosen to be 100.  For each box, 1,000 independently seeded stochastic 
trajectories were calculated until completion.  Looking at the $SZR$
results in the top row, we can see that the dynamics is fairly independent of
population size once the population size gets above around 100 individuals.
The population dependence for lower population sizes is an effect of the
stochasticity.  We can clearly see a transition in the susceptible population
near $\alpha = 1$ corresponding to where our continuous dynamics would show a
sharp boundary.  Here the boundary is blurred, again due to the stochasticity.
The final dead zombie population $R$ remains small for all values of $\alpha$;
for extremely virulent zombies $\alpha \ll 1$, very few will be killed by the
humans before all of the humans are converted, while in the other extreme few
zombies are created so there are few to be killed.

Contrast these results with the density dependent $SIR$ dynamics shown in the
second row.  There can be no infected individuals left in the end, so only the
fraction of $S$ and $R$ in the final state are shown.  The two transitions in $SIR$
couple differently to the population of infected and susceptible.  While our
nondimensionalized $SZR$ model has $Z' = (1-\alpha)SZ/N$, our nondimensionlized
$SIR$ has $I' = ( S/N - \mu ) I$.  This creates a very strong population
dependence.  The transition observed in the $S$ population is largely independent
of $\mu$, except on the very small end.  When we move to inhomogeneous
population lattices this means that for the density dependent $SIR$ model, the
most important parameter governing whether a particular site has a break-out
infection is the population of that site on the lattice.

\section{Critical Behavior of Lattice Model}
\label{sec:RG}

Until now, we have considered fully connected, well-mixed populations, where any infected
individual can infect any susceptible individual with equal probability.  But surely, a zombie in New
York cannot bite someone in Los Angeles. Investigation of the spatial spread of
infectious diseases is an important application of {\em network science}; social
diseases spread among intimate contacts, Ebola spreads by personal contact in a
network of care-givers, influenza can be spread by direct contact, through the
air or by hand-to-mouth, hand-to-eye or hand-to-nose contact after exposure to
a contaminated surface. For most diseases, `long bonds' dominate the
propagation to distant sites~\cite{smallworld}; airplane flights take Ebola to new continents.
Zombies do not fly airplanes, so our model is closer in spirit to the spread of
certain agricultural infestations, where the disease spreads across a lattice
of sites along the two-dimensional surface of the Earth (although not in those cases 
where pathogens are transported long distances by atmospheric currents).

To begin, we will consider a two-dimensional square lattice, where each site contains a
single individual.  Each individual is allowed to be in one of three states: $S,
Z$, or $R$.  The infection spreads through nearest neighbor bonds only.  That is,
a zombie can bite or be killed by any susceptible individuals in each of the
four neighboring sites. 

To make direct contact with our zombie model, the rate at which an susceptible
cell is bitten is given by $\beta Z$ where $Z$ is the number of zombie
neighbors (since $S$ is one), and the rate at which a zombie site is killed is
$\kappa S$ where $S$ is the number of susceptible neighbors.  

Because all state transitions in the $SZR$ model depend only on $Z$--$S$ contacts, 
for computational efficiency, we need only maintain a queue of all $Z$--$S$ bonds,
that is connections along which a human and zombie can interact.  At each step
of the simulation, one of these $Z$--$S$ bonds is chosen at random, and with
probability $\beta/(\beta+\kappa) = 1/(1+\alpha)$, the human is bitten, marking
it as a zombie. We can then query its neighbors, and for all of them that are
human, we can add a $Z$--$S$ link to our queue.  With probability $\kappa / (\beta +
\kappa) = \alpha/(1+\alpha)$ the zombie is killed, removing any of its links to
neighboring humans from the queue.  This process matches the stochastic
dynamics of our zombie model operating on the lattice.

Simulating zombie outbreaks on fixed lattices, there is qualitatively different
behavior for small $\alpha$ and large $\alpha$.  When $\alpha$ is large, the
zombies do not spread very far, always being defeated by their neighboring
humans.  When $\alpha$ is very small, the zombies seem to grow until they
infect the entire lattice.  This suggests evidence of a phase transition.
Technically, the presence of a phase transition would mean that if we could
simulate our model on an infinite lattice, there should be some critical
$\alpha$ ($\alpha_c$), above which any outbreak will necessarily terminate. 
Below the critical value, there is the possibility (assuming the infection does not die out) of having
the infection grow without bound, infecting a finite fraction of individuals in the limit that
the lattice size becomes infinite.  The $SIR$ model has been demonstrated to undergo such a phase 
transition, and we expect the zombie model does as well.

The study of \emph{critical phenomena} includes a series of techniques and
analyses that enable us to study the properties of phase
transitions even on finite lattices.  A major theme of critical phase
transitions is the importance of {\em critical points} -- where a system
is tuned (here by varying $\alpha$) to a value separating qualitatively
different behaviors (here separating low-infectivity transient zombie
infestations from a potentially world-spanning epidemic). At critical
points, the system can show {\em scale free behavior}; there is no natural
length scale to the dynamics, and various physical parameters will usually
be governed by {\em power laws} (see below).

\begin{figure}[htbp] 
    \includegraphics[width=\linewidth]{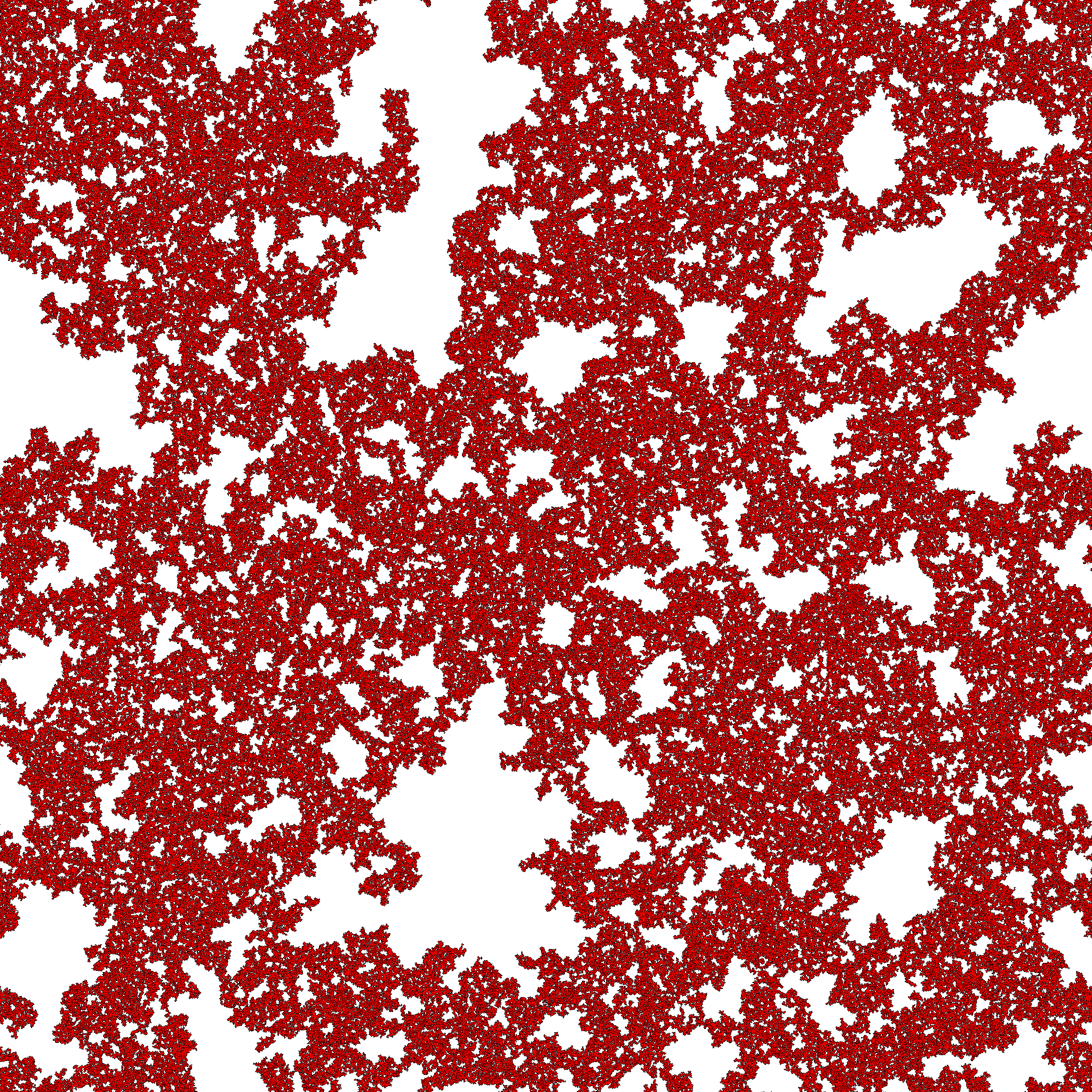}

    \caption{Example cluster resulting from the single population per site square lattice zombie model with periodic boundary conditions near the critical point $\alpha_c = \critpoint$ on a lattice of size
        $2048\times 2048$.
    }
    \label{fig:clust} 
\end{figure}

With $\alpha$ chosen to be precisely at the critical
value, we indeed see a giant component with fractal structure
(Fig.~\ref{fig:clust}). Note that there are
holes (surviving pockets of humans) of all sizes in the figure. This 
reflects the proximity to the threshold: the battle between zombies and humans
is so evenly matched, that one gets an {\em emergent scale invariance} in
the survival patterns. This is in keeping with studies of the $SIR$ model, which shows a similar critical behavior
and phase transition~\cite{grassberger}.

Systems near critical points with this kind of scale invariance fall into
{\em universality classes}. Different systems (say, a real disease outbreak
and a simple computational model) can in many ways act precisely the same
on large scales near their transitions (allowing us to predict behavior
without knowing the details of zombie-human (anti)social interactions).
The $SIR$ model on a two-dimensional lattice with a single person per site 
falls into the percolation universality class~\cite{cardy}, though details of 
its cluster growth can differ~\cite{tome}. 
Given that the $SZR$ model has two second order couplings, it is
of interest whether it falls into the same percolation universality class.

To extract the scaling behavior of our zombie infestation, we study the 
distribution $P(s, \alpha)$, the probability that a single zombie
will generate an outbreak of size $s$ at inverse virulence $\alpha$. 
(An outbreak will be a fractal cluster in two dimensions, with ragged
boundaries if it dies out before reaching the entire world.)
At $\alpha = \alpha_c$ where the zombies and humans are equally matched,
we have an emergent scale invariance. A large outbreak will appear to almost
stop several times -- it can be viewed as a sequence of medium-sized outbreaks
triggering one another. Medium-sized outbreaks are composed of small outbreaks,
which are in turn composed of tiny outbreaks. At threshold, each of these
scales (large, medium, small) is related to the lower scale (medium, small,
tiny) in the same fashion. Let us oversimplify to say that at criticality
an outbreak of size $3 s$ is formed by what would have been three smaller
outbreaks of size $s$ which happened to trigger one another, and these in
turn are formed by what would have been three outbreaks of size $s/3$.
If the probabilities and form of this mutual triggering is the same at each
scale, then it would not surprise us that many properties of the outbreaks
would be the same, after rescaling the sizes by a factor of three. In
particular, we expect at the critical point to find the probabilities
of outbreaks of size $s$ to be related to the probabilities at size $s/3$
by some factor $f$:
\begin{equation}
P(s, \alpha_c) = f P(s/3, \alpha_c).
\end{equation}
This formula implies that $P(s, \alpha_c) \propto s^{-\tau}$, with
$\tau = \log(1/f)/\log(3)$. The distribution of epidemic infection rates
is a power law.

\begin{figure}[htbp] 
    \includegraphics[width=\linewidth]{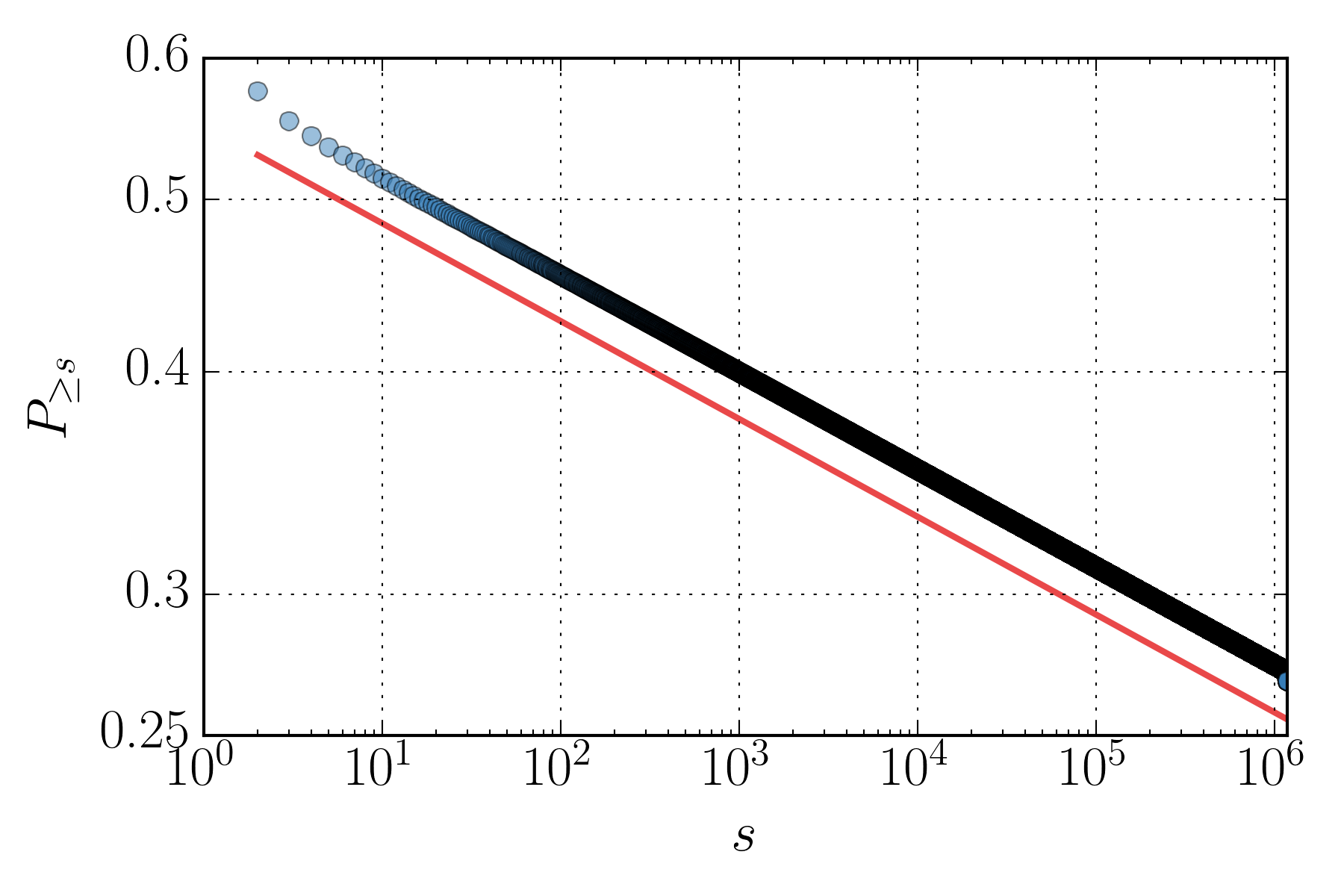}
    \includegraphics[width=\linewidth]{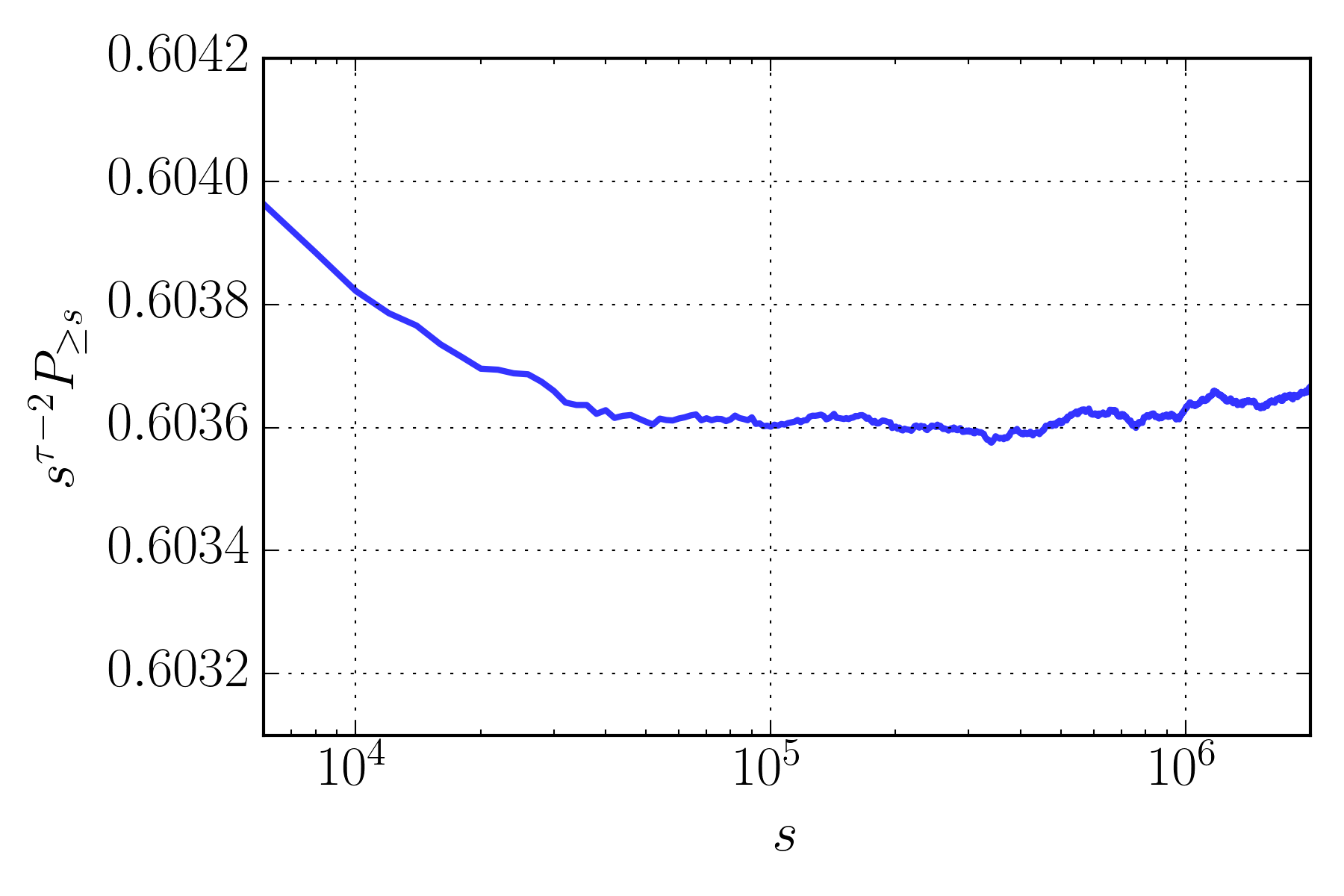}

    \caption{The cumulative distribution of epidemic sizes for
the two dimensional zombie model near the critical virulence.
The critical point found was $\alpha_c = \critpoint$. The top plot
shows the probability of a site being in a cluster of
at least $s$ in size ($P_{\geq s}$). 
The fact that it forms a straight
line on a log-log plot indicates that $P_{\geq s}$ is a power law, and 
the slope is $2-\tau$. 
For comparison, the red (color online) line shows
the powerlaw corresponding to the percolation critical exponent: $\tau = 187/91$. 
The bottom plot shows the same data times $s^{\tau-2}$
using the exponent from percolation theory.
The plot is very nearly flat suggesting the percolation exponent accurately
describes the zombie model.
}

    \label{fig:scaling} 
\end{figure}

Figure~\ref{fig:scaling} shows a thorough test of this dependence for our
zombie model, following a procedure akin to that of reference~\cite{tome}.
We simulated a
zombie outbreak on a two-dimensional lattice with periodic boundary conditions
starting with a single zombie.  
With the outbreak sizes
following a power law distribution, the probability that a site belongs to a
cluster of size $n_s$ is $P_s = s n_s$, so that at the critical point $P_s \sim
s^{1-\tau}$.  
Integrating from $s$ to $\infty$, the probability that a point
belongs to a cluster of at least $s$ in size ($P_{\geq s}$) should at the critical point
itself follow a powerlaw: $P_{\geq s} \sim s^{2-\tau}$.  To find our critical point $\alpha_c$, we
ran many simulations until our integrated cluster size distribution followed a power law, using the interpolation methods of reference~\cite{tome} to get a precise estimate of the critical point.

For zombies on a two dimensional lattice, this critical point occurs at
$\alpha_c = \critpoint$, the resulting integrated cluster size distribution
is shown at the top of Fig.~\ref{fig:scaling}.
Percolation theory predicts $\tau = 187/91$ in two dimensions, and we
test that prediction in the bottom part of Fig.~\ref{fig:scaling}. Here, if we were precisely at the
critical point and the $SZR$ model is in the percolation universality class, with infinite statistics we would have asymptotically a perfectly straight line.  Notice the small 
vertical scale: our fractional fluctuations are less than 0.1\%, while our experimental results vary over several order of magnitude.
The clear agreement convincingly shows that the zombie model on the two dimensional 
lattice is in the percolation university class.

\begin{figure}[htbp] 
    \includegraphics[width=0.9\linewidth]{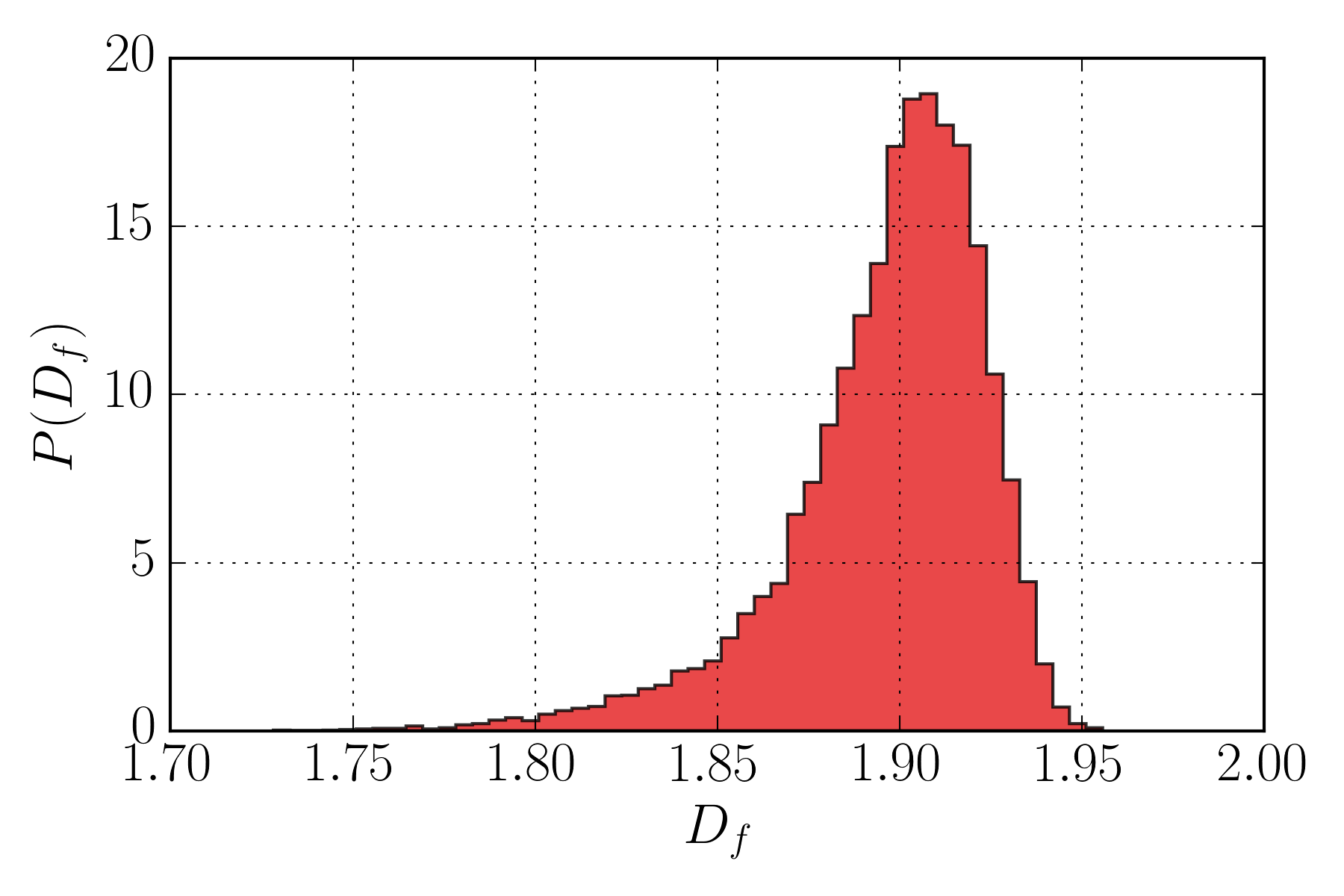}

    \caption{ A histogram of the observed fractal dimension of the zombie
        epidemic clusters as measured by box counting.  These give a measured
        value of $D=1.8946(14)$, consist with the exact percolation fractal
        dimension of $D=91/48 = 1.895833$.  }

    \label{fig:box} 
\end{figure}

As an additional check, we computed the fractal dimension of our clusters near the critical point
using box counting, a distribution for which is shown in Figure~\ref{fig:box}. We find a fractal dimension
$D = 1.8946(14)$, compared to the exact percolation value of $D= 91/48 = 1.895833$.

Why did we need such an exhaustive test (many decades of scaling, many
digits in our estimate of $\alpha_c$)? On the
one hand, a much smaller simulation could have told us that there was
emergent scale invariance and fractal behavior near the transition; one
or two decades of scaling should be convincing. But it turns out that there
are multiple different universality classes for this kind of invasion process,
and their exponents $\tau$ and $D$ are rather similar. And a small error
in $\alpha_c$ can produce large shifts in the resulting fits for $\tau$ and
$D$ -- demanding efficient programming and fast computers to achieve a
definitive answer.

We conclude that the single person per site zombie infestation, near the critical virulence, 
will on long length scales develop spatial infestation patterns that are
well described by two-dimensional percolation theory.

\section{US Scale Simulation of Zombie Outbreak} 
\label{sec:US}

Having explored the general behavior of the zombie model analytically,
stochastically and on homogeneous single person lattices, we are prepared to
simulate a full scale zombie outbreak.

\subsection{Inhomogeneous Population Lattice}

\begin{figure}[htbp]
    \includegraphics[width=\linewidth]{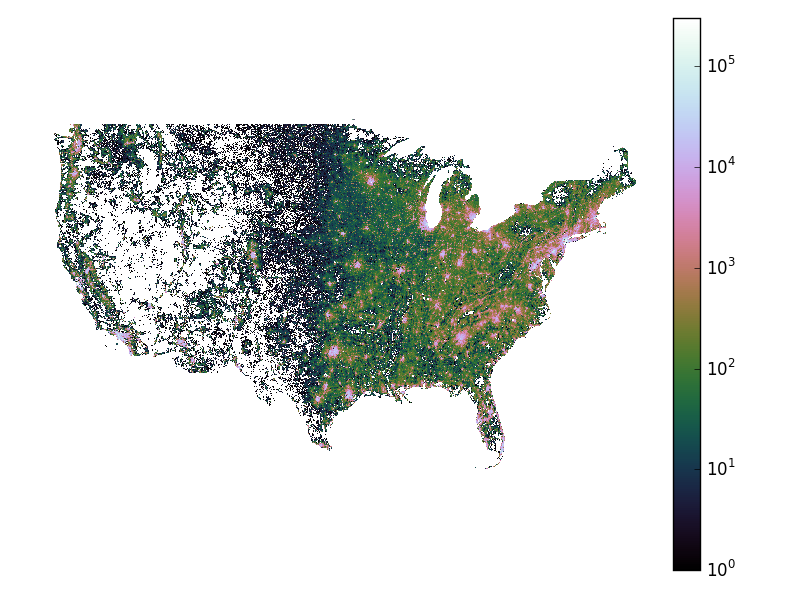}
    \includegraphics[width=\linewidth]{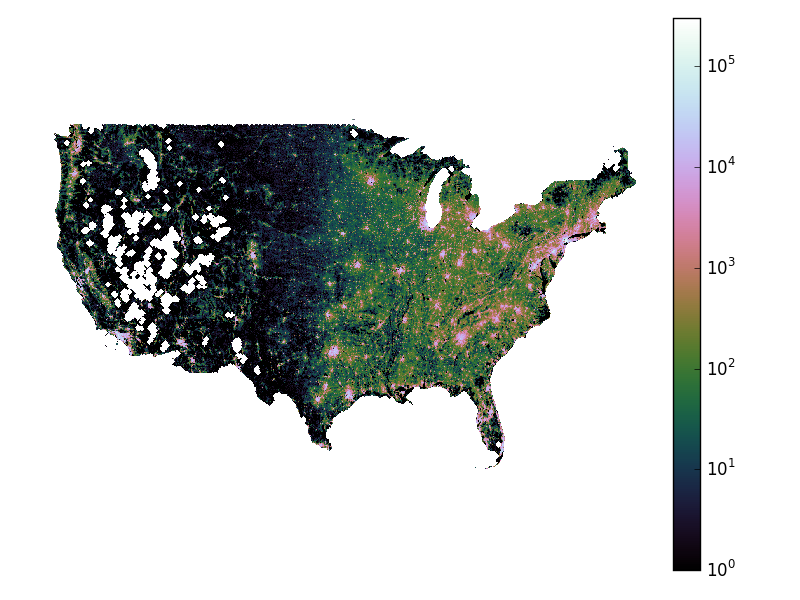}

    \caption{A $1500 \times 900$ grid of the 2010 US Census Data.  The above figure
    gives the raw results.  Notice the multitude of squares with no people in
them in the Western United States. The bottom figure shows the resulting map
after 6 steps of binary closing added to the original population.}

    \label{fig:population}
\end{figure}

We will attempt to simulate a zombie outbreak occurring in the United States.
This will be similar to our lattice simulation, but with an inhomogeneous
population lattice.  We based our lattice on code available for creating a
``dot map" based off the 2010 US Census
data \footnote{\url{https://github.com/meetar/dotmap}}.  The 2010 Census
released census block level data, detailing the location and population of 11,155,486
different blocks in the United States.  To cast these blocks down to a square
grid, we assigned each of the 306,675,005 reported individuals a random
location inside their corresponding census block, then gridded the population
into a $1500 \times 900$ grid based on latitude and longitude coordinates. The resulting
population lattice can be seen in the top half of Figure~\ref{fig:population}.
You will see the presence of many empty grids, especially throughout the
western United States. This disconnects the east and west coasts in a clearly
artificial pattern -- our zombies in practice will gradually wander through
the empty grid points.
To add in lattice connectivity, we did six iterations
of binary closing (an image processing technique)
on the population lattice and added it to the original.  The
effect was to add a single person to many vacant sites, taking our total
population up to 307,407,336.  The resulting population map is shown in the
bottom half of Figure~ \ref{fig:population}.  This grid size corresponds to
roughly 3 km square boxes.  The most populated grid site is downtown New York
City, with 299,616 individuals.  The mean population of the occupied grid sites
is 420, the median population of an occupied site is 13.

\subsection{Augmented Model}

In order to more `realistically' simulate a zombie outbreak, we made two
additions to our simplified $SZR$ model.  The first was to add a latent state
$E$ (Exposed). The second was to introduce motion for the zombies. Considered as a system
of differential equations, we now have:
\begin{align}
    \dot S_{i} &= -\beta S_{i} Z_{i} \\
    \dot E_{i} &= -\nu E_{i} \\
    \dot Z_{i} &= \nu E_{i} - \kappa S_{i} Z_{i} \\
    \dot R_{i} &= \kappa S_{i} Z_{i} \\
    \dot Z_{i} &= \mu \sum_{\langle j \rangle} Z_{j} - \mu Z_{i}
\end{align}
or as a set of reactions:
\begin{align}
    ( S_i, E_i ) &\xrightarrow{\makebox[1cm]{$\beta S_i Z_i$}} (S_i -1 , E_i+1) \\
    ( Z_i, E_i ) &\xrightarrow{\makebox[1cm]{$\nu E_i$}} ( Z_i+1, E_i-1) \\
    ( Z_i, R_i ) &\xrightarrow{\makebox[1cm]{$\kappa S_i Z_i$}} ( Z_i -1 , R_i+1) \\
    \langle i\, j \rangle : ( Z_i, Z_j ) &\xrightarrow{\makebox[1cm]{$\mu Z_i$} } ( Z_i -1 , Z_j+1)\ .
\end{align}
Here $i$ denotes a particular site on our lattice.  $\langle j \rangle$ denotes
a sum over nearest neighbor sites, $\langle i \, j \rangle$ denotes that $i$
and $j$ are nearest neighbors. In this model, zombies and humans only interact
if they are at the same site, but the zombies diffuse on the lattice, being
allowed to move to a neighboring site with probability proportional to their
population and some diffusion constant ($\mu$).  We assume that the humans do not move,
not only for computational efficiency, but because, as we will see, the zombie outbreaks
tend to happen rather quickly, and we expect large transportation networks to shut down in 
the first days, pinning most people to their homes.  The addition of a latent state
coincides with the common depiction that once a human has been bitten, it
typically takes some amount of time before they die and reanimate as a zombie.
If a human is bitten, they transition to the $E$ state, where at some constant
rate ($\nu$) they convert into the zombie state.

To choose our parameters we tried to reflect common depictions of zombies in
movies.  The work of Witkowski and Blais~\cite{bayesian} performed a
Bayesian fit of a very similar $SZR$ model to two films, \emph{Night of the
Living Dead}, and \emph{Shawn of the Dead}.  In both cases, the observed
$\alpha$ was very close to 0.8. This means that the zombies in the films are
1.25 times more effective at biting humans than the humans are at killing the
zombies. We will adopt this value for our simulation. For our latent state, we
adopt a value close to that reported for \emph{Shawn of the Dead}, namely a
half-life of 30 minutes. To set our movement parameter, we estimate that
zombies move at around 1 ft/sec.  To estimate the rate at which the zombies
will transition from one cell to the next, we assume that the zombies behave
like a random gas inside the cell, so that the probability that a zombie will
cross a cell boundary is roughly $\frac 14 \frac{Z}{L^2} L v \Delta t$, that
is, one-fourth of the zombies within $v \Delta t$ of the edge will move across
that edge in a small amount of time.  This suggests a value of $\mu$ of $0.0914
\text{ /hr}$.  This corresponds to an average time between transitions of
around 11 hours, which for a zombie stumbling around a 3 km block agrees with
our intuitions.  Finally, to set a rate for our bite parameter, we similarly
assume that the zombies are undergoing random motion inside the cell at 1
ft/sec, and they interact with a human anytime they come within 100 feet.  We
can then estimate the rate at which humans and zombies will interact as $S Z
\frac{ R v \Delta t}{ L^2 }$, which corresponds to a choice of $\beta$ of around
$3.6 \times 10^{-3} \text{ /hr}$.  Another way to make sense of these parameter
choices is to ask how many susceptible individuals must be in a cell before a
single zombie has a higher rate for biting a human than transitioning to a
neighboring cell.  For our choice of parameters, this gives 
\begin{align} N
    \beta = 4 \mu \implies N \sim 102 \ .  
\end{align} 
This corresponds to a low
population density of \mbox{$\sim11\text{ people/km$^2$}$}, again agreeing with our
intuition.  All of our parameter choices are summarized in Table
\ref{tab:param}.

\begin{table}
    \begin{center}
    \begin{tabular}{r|l}
        $\beta$ & $3.6 \times 10^{-3} \text{ /hr/person}$ \\
        \hline
        $\alpha$ & 0.8 \\
        \hline
        $\kappa$ & $\alpha \beta$ \\
        \hline
        $\eta$ & $2 \text{ /hr}$ \\
        \hline
        $\mu$ & $0.0914 \text{ /hr}$
    \end{tabular}
    \end{center}

    \caption{The parameters chosen for our US-scale simulations of a zombie outbreak.  These parameters
        were chosen to correspond with standard depictions of zombies and simple physical estimations explained
    in the main text.}

    \label{tab:param}
\end{table}

\subsection{Simulation Details}

To effectively simulate an outbreak at this scale, we employed the Next
Reaction Method of~\cite{bettergillespie}.  We maintained a priority queue of
all possible reactions, assigning each the time at which the reaction would
take place, an exponentially distributed random number with scale set by the
rate for the reaction.  At each time step of the simulation, we popped the next
reaction off of the queue, and updated the state of the relevant squares on our
grid.  Whenever population counts changed, we of course needed to update the
times for the reactions that depend on those population counts.  This method
remained efficient for simulating the entire US. However, at late times
a large amount of simulation time was spent simulating the diffusion of
the zombies back and
forth between highly populated states.  We could have achieved additional
computational efficiency by adopting the time dependent propensity function
approach of Fu et al.~\cite{fu}.

\subsection{Results}

\begin{figure}[htbp]
    \begin{center}
    \includegraphics[width=\linewidth]{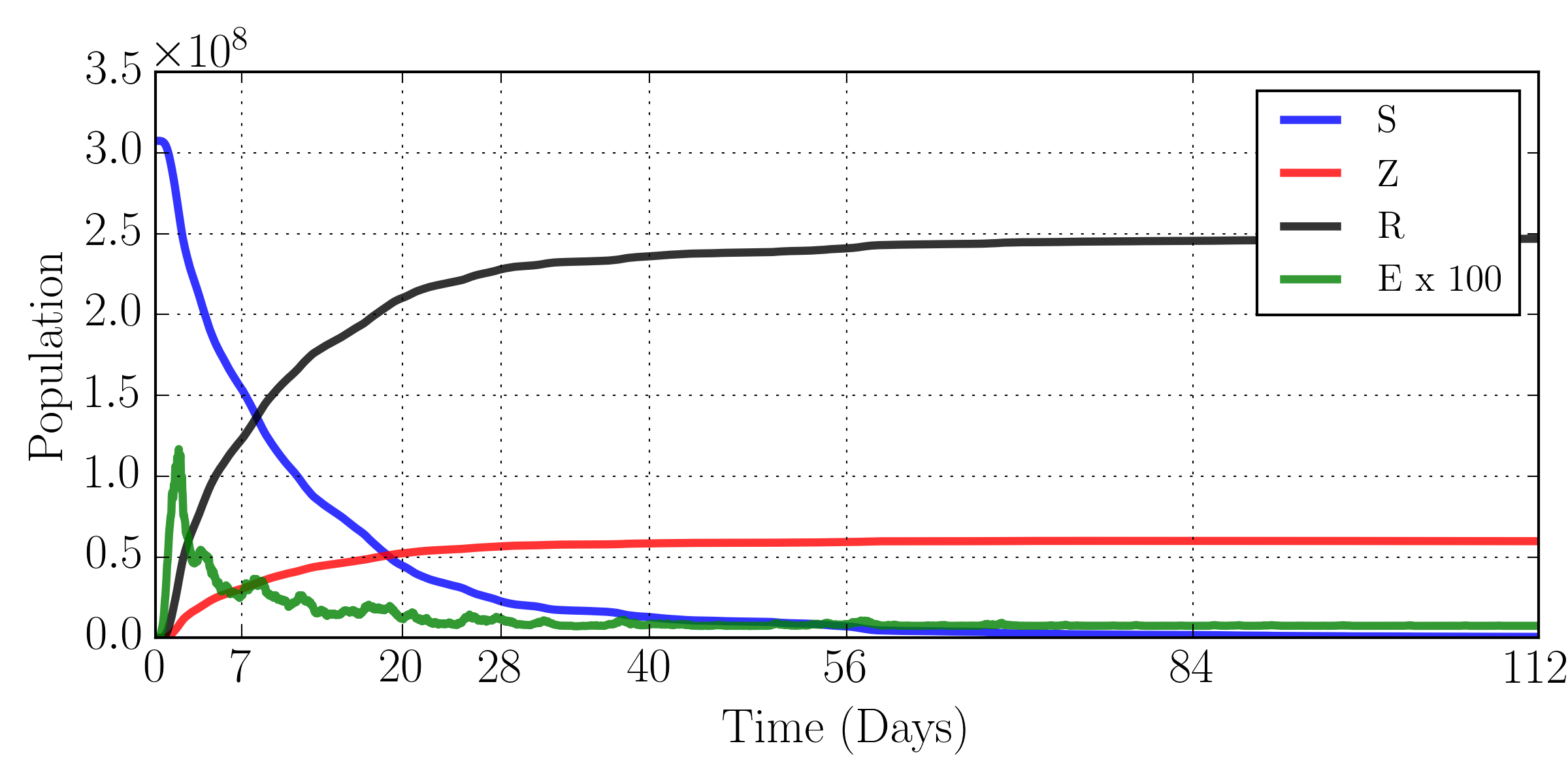}
    \end{center}

    \caption{The $S$ (blue), $Z$ (red), $R$ (black), and $E$ (green)
    populations as a function of time for a full scale zombie outbreak in the
continental United States starting with one in every million people infected
(color online). The exposed population ($E$) has been magnified by a factor of 100. }

    \label{fig:overall}
\end{figure}

\begin{figure}[htbp]
    \begin{center}
    \subfloat[1 Day]{\includegraphics[width=0.5\linewidth]{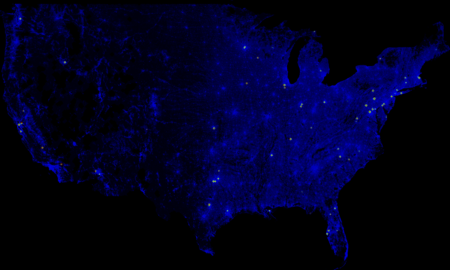}}
    \subfloat[2 Days]{\includegraphics[width=0.5\linewidth]{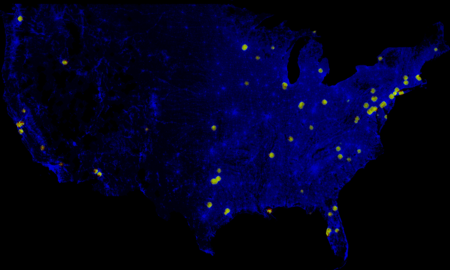}}
\\  
    \subfloat[1 Week]{\includegraphics[width=0.5\linewidth]{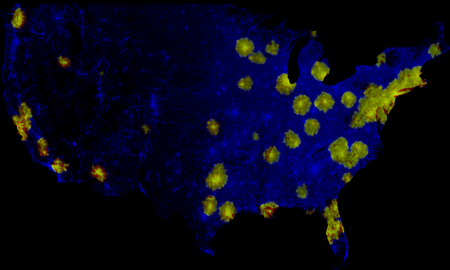}}
    \subfloat[2 Weeks]{\includegraphics[width=0.5\linewidth]{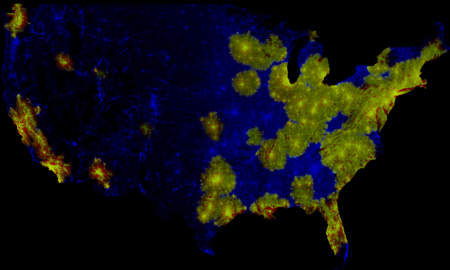}}
\\
    \subfloat[3 Weeks]{\includegraphics[width=0.5\linewidth]{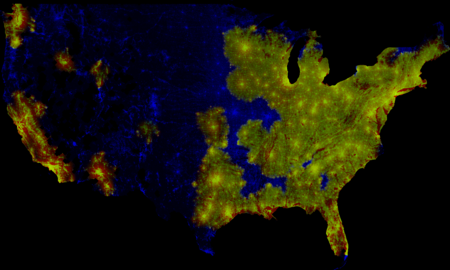}}
    \subfloat[4 Weeks]{\includegraphics[width=0.5\linewidth]{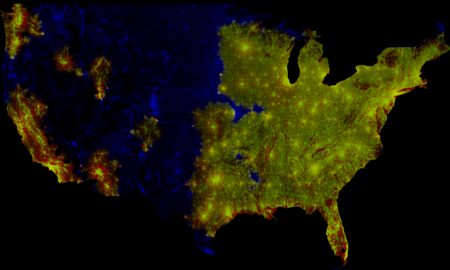}}
\\
    \subfloat[2 Months]{\includegraphics[width=0.5\linewidth]{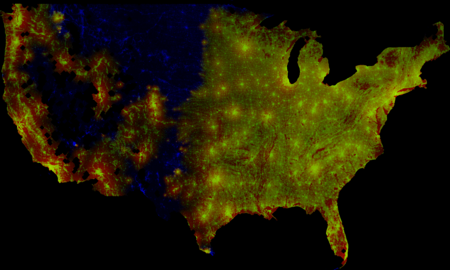}}
    \subfloat[4 Months]{\includegraphics[width=0.5\linewidth]{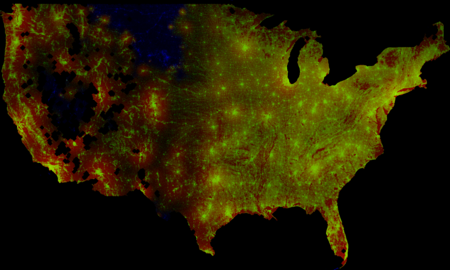}}
    \end{center}

    \caption{Simulation of a zombie outbreak in the continental United States.
Initially one in every million individuals was infected at random.  Results are
shown above at (a) one day, (b) two days, (c) one week, (d) two weeks, (e)
three weeks, (f) four weeks, and (g) two months after the outbreak begins.
Shown here are the population of susceptible individuals ($S$) in blue, scaled
logarithmically, zombies in red and removed in green (color online).  All three channels are
superimposed. A movie version of this outbreak is available in the supplemental materials online. }

    \label{fig:movie}
\end{figure}

With the simulation in place, we are now in a position to simulate a full scale
zombie outbreak.  We first consider an outbreak that began with one in every
million individuals starting in the Exposed ($E$) state in the United States.
For a single instance the overall populations are shown in Figure
\ref{fig:overall}.  This looks similar to the analytical outbreaks we saw in
Figure~\ref{fig:analytic_example}, but with a steeper rate of initial infection
and some slight perturbations to the curves.  The total population curves
however hide most of the interesting features.  In Figure~\ref{fig:movie} we
attempt to give a sense of how this outbreak evolves, showing the state of the
United States at various times after the outbreak begins. 

As you can see, for the parameters we chose, most of the United States
population has been turned into zombies by the first week, while the geographic
map does not necessarily seem all that compelling. In the early stages of the
outbreak, while the population is roughly homogeneous, the zombie plague spreads
out in roughly uniform circles, where the speed of the infection is tied to the
local population density. Infestations on the coasts, with their higher
population density, have spread farther than those near the center of
the country. After several
weeks, the map exhibits stronger anisotropy, as we spread
over larger geographical areas and the zombie front is influenced by large
inhomogeneities in population density. After four weeks, much of the United
States has fallen, but 
it takes a very long time for the zombies to diffuse and capture the
remaining portions of the United States. Even four months in, remote areas of
Montana and Nevada remain zombie free.

\begin{figure}[htbp]
    \includegraphics[width=\linewidth]{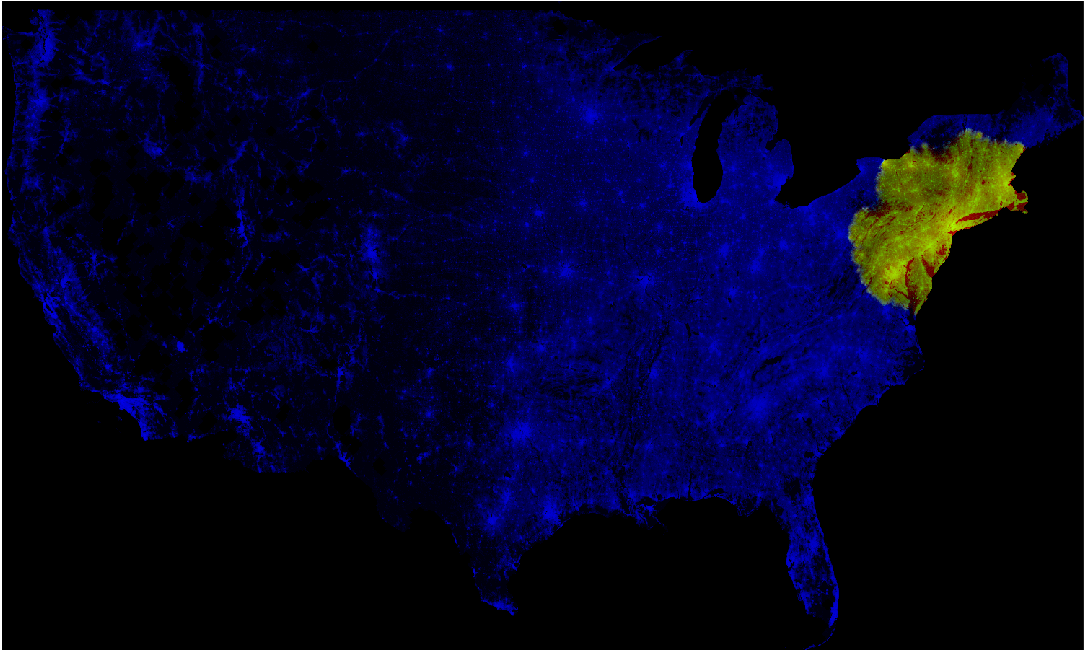}

    \caption{ Status of the United States 28 days after an outbreak that
        started in New York City. Here blue represents humans, red represents
        zombies and green represents dead zombies (color online).  The three color channels
        have been laid on top of one another.  }

    \label{fig:nyc} 
\end{figure}

To investigate the geographical characteristics of the outbreak, we must move
beyond a single instance of an outbreak and study how different regions are
affected in an ensemble of outbreaks. If it takes a month to develop
and distribute an effective vaccine (or an effective strategy for zombie
decapitation), what regions should one locate the zombie-fighting headquarters?
We ran 7,000 different 28-day zombie
outbreaks in the continental United States starting with a single individual.
A single instance of one of these outbreaks originating in New York City
is shown in Figure~\ref{fig:nyc}.

\begin{figure*}[htbp]
    \includegraphics[width=\linewidth]{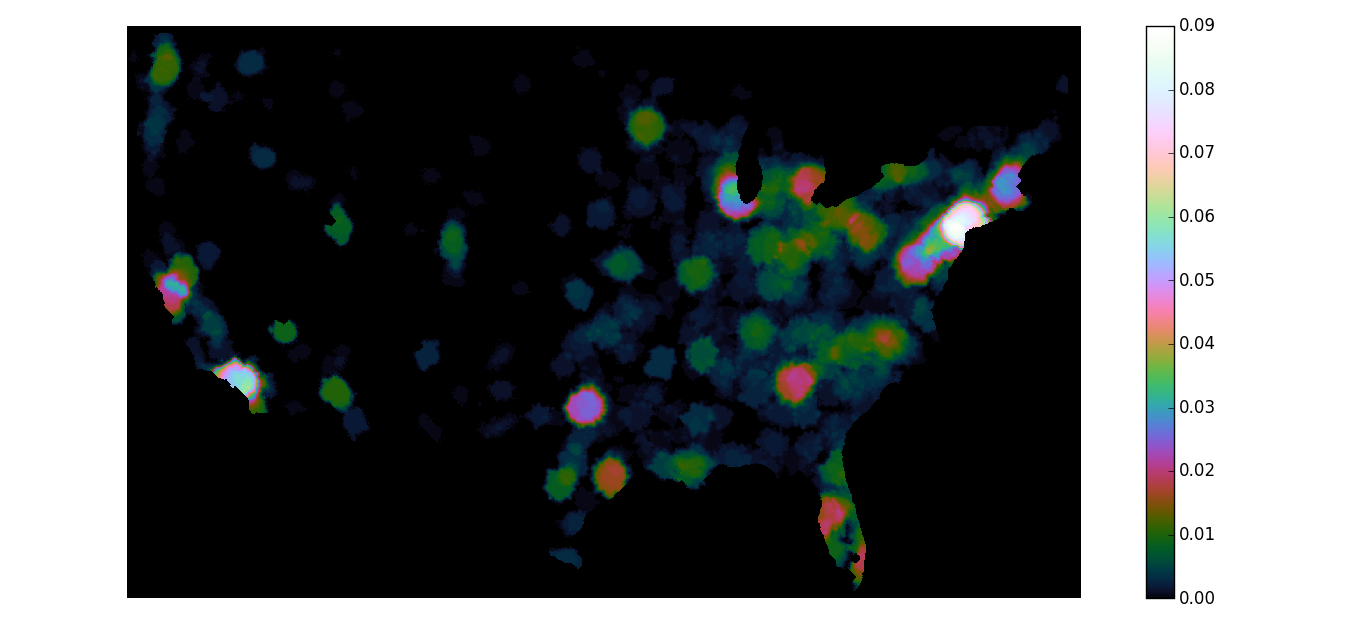}
    \includegraphics[width=\linewidth]{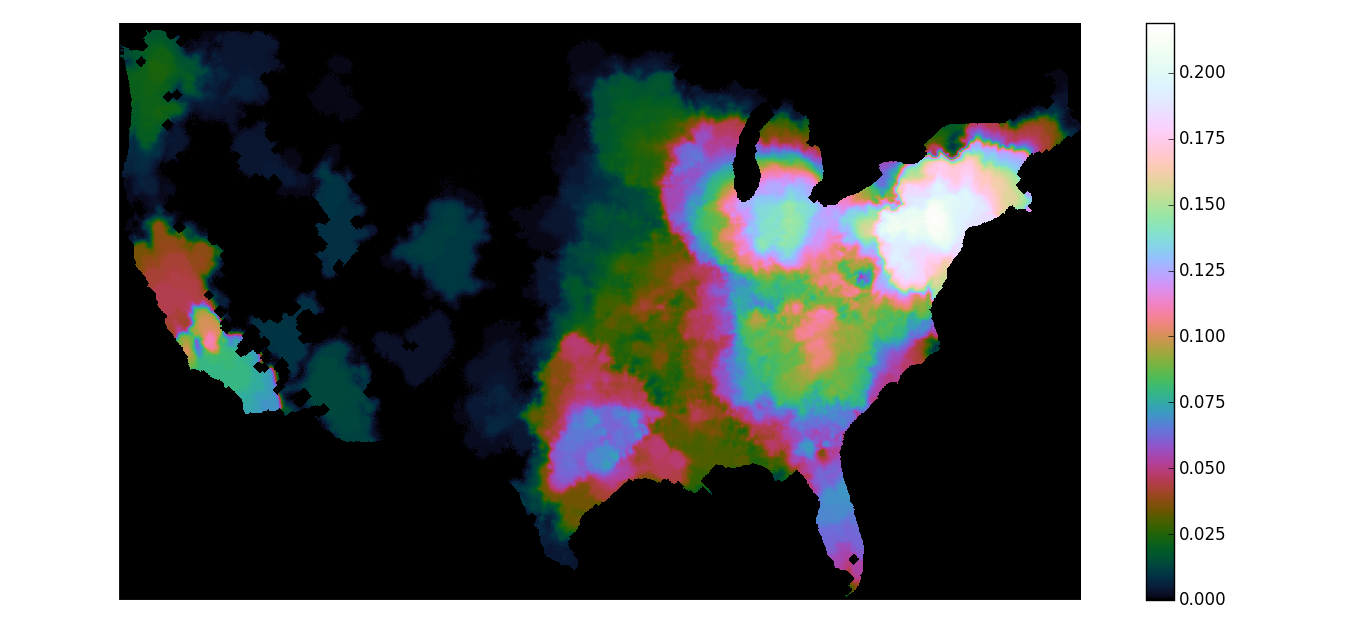}

    \caption{ Average survivability from US scale runs. In both cases, the plot
    shows the probability of being infected in that square after an epidemic
that originates from a single infected individual chosen at random from the
total population.  The top figure is the probability of being infected after 7
days, while the bottom plot is after 28 days. In total, this represents 7,000
simulated runs starting from a single individual.  The top plot represents the
1,467 outbreaks that lasted at least 7 days, the bottom plot represents 1,458
outbreaks that lasted at least 28 days.  }

    \label{fig:US_survive} 
\end{figure*}

By averaging over all of these runs, we can start to build a zombie
susceptibility map, as shown in Figure~\ref{fig:US_survive}.  In the top plot,
we show the probability that the given cell is overrun by zombies after seven
days.  Here you can clearly see that there are certain regions -- those
surrounding populous metropolitan areas -- that are at a greater risk.  This is
partly because those regions have lots of individuals who could potential
serve as patient zero, and partly due to the rapid spread of zombies in
those areas.  In the bottom plot, we plot the probability that the
cell is overrun, but at the 28 day mark.

After 28 days, it is not the largest metropolitan areas that suffer the
greatest risk, but the regions located between large metropolitan areas. 
For instance, in California it is the region near
Bakersfield in the San Joaquin Valley that is at the greatest risk as this area
will be overrun by zombies whether they originate in the San Francisco area or
the Los Angeles / San Diego area. The area with the greatest one month zombie
risk is north eastern Pennsylvania, itself being susceptible to outbreaks
originating in any of the large metropolitan areas on the east coast.

\section{Conclusion}

Zombies offer a fun framework for introducing many modern
concepts from epidemiology and critical phenomena.
We have described and analyzed various zombie models, from one describing 
deterministic dynamics in a well-mixed system to a full scale US epidemic.
We have given a closed form analytical solution to the well-mixed dynamic
differential equation model.  We compared the stochastic dynamics to a
comparable density-dependent $SIR$ model. We investigated the critical behavior of the single person per site two-dimensional square lattice zombie
model and demonstrated it is in the percolation universality class.  We ran
full scale simulations of a zombie epidemic, incorporating each human
in the continental United States, and discussed the geographical implications
for survival.

While this work is predicated on a fictional infestation, one might
ask whether there are any phenomena in the real world that behave 
in a manner similar to our modeled zombie outbreaks.  As noted,
the $SZR$ model requires that susceptible hosts directly participate
in the removal of zombie hosts from the infectious population, leading
to runaway outbreaks as susceptible hosts are depleted.  One might
imagine a similar phenomenon for infectious diseases that require
medical intervention to be suppressed; as medical personnel themselves
become infected (as has sadly happened to a considerable degree during
the recent Ebola outbreak in West Africa), they become less able to
stem the spread of infection.  (Medical personnel, however, represent only 
a small fraction of all susceptible hosts, so a refinement to an $SZR$-type model
would be required to account for this.)  One might also imagine $SZR$-like dynamics in 
the spread of ideas and opinions: a person spreading a controversial opinion in a
population, for example, might be able to sway some converts, but is also likely to meet 
resistance and counter-arguments, which act to reduce infectivity and
perhaps ultimately stop the spread. 

We hope our systematic treatment of an imaginary disease will provide a 
useful and inspiring teaser for the exciting fields of statistical mechanics,
network science, and epidemiology.

\section{Acknowledgments}

We acknowledge NSF IIS--1247696 and Cornell University for support of this research, and thank Paul Ginsparg for useful references and conversations.

\bibliographystyle{plain}
\bibliography{zombieref}

\end{document}